\documentclass{jnmp03c}

\begin{document}
\setcounter{page}{38}

\renewcommand{\evenhead}{H S Dhillon, F V Kusmartsev and  K E K\"urten}
\renewcommand{\oddhead}{Fractal and Chaotic Solutions of the Discrete Nonlinear
Schr\"odinger Equation}

\thispagestyle{empty}

\FistPageHead{1}{\pageref{dhillon-firstpage}--\pageref{dhillon-lastpage}}{Letter}

\copyrightnote{2001}{H S Dhillon, F V Kusmartsev and  K E
K\"urten}

\Name{Fractal and Chaotic Solutions of the Discrete Nonlinear
Schr\"odinger Equation in Classical\\ and Quantum Systems}\label{dhillon-firstpage}

\Author{H S DHILLON~$^{\dag^1}$,
F V KUSMARTSEV~$^{\dag^1\, \dag^2}$ and  K E K\"URTEN~$^{\dag^3}$}

\Adress{$^{\dag^1}$~Department of Physics,
Loughborough University, LE11 3TU, UK\\[2mm]
$^{\dag^2}$~Landau Institute, Moscow, Russia\\[2mm]
$^{\dag^3}$~Institut f\"ur Experimentalphysik, Universit\"at Wien, Austria}

\Date{Received May 8, 2000; Accepted August 30, 2000}

\begin{abstract}
\noindent
We discuss stationary solutions of the discrete nonlinear Schr\"odinger
equation (DNSE) with a potential of the $\phi^{4}$ type which is
generically applicable to several quantum spin, electron and classical
lattice systems.  We show that there may arise chaotic spatial
structures in the form of incommensurate or irregular quantum states.
As a first (ty\-pi\-cal) example we consider a single electron which is
strongly coupled with phonons on a $1D$ chain of atoms --- the
(Rashba)--Holstein polaron model.  In the adiabatic approximation
this system is conventionally described by the DNSE.
Another relevant example is that of superconducting states
in layered superconductors described by the same
DNSE.  Amongst many other applications the typical example for a classical
lattice is a system of coupled nonlinear oscillators.
We present the exact energy spectrum of this model in the strong
coupling limit and the corresponding wave function.
Using this as a starting point we go on to calculate the wave
function for moderate coupling and find that the
energy eigenvalue of these structures of the wave function is in
exquisite agreement with the exact strong coupling result.
This procedure allows us to obtain (numerically) exact solutions
of the DNSE directly.  When applied to our
typical example we find that the wave function
of an electron on a deformable lattice (and other quantum or classical
discrete systems) may exhibit incommensurate
and irregular structures.  These states are analogous to the
periodic, quasiperiodic and chaotic structures found in
classical chaotic dynamics.
\end{abstract}

\section{Introduction}

Chaos is an important branch of nonlinear dynamics because
chaotic behaviour seems to be universal \cite{Hilborn}.  It is
present in mechanical oscillators, electrical circuits, lasers,
nonlinear optical systems, chemical reactions, nerve cells, heated
fluids and weather systems.  Even more importantly, this chaotic
behaviour shows qualitative and quantitative universal features
which are independent of the details of the particular system
and corresponds to a disappearance of periodic trajectories.

It is commonly believed that in quantum systems chaotic structures
cannot arise.  Ho\-we\-ver, the question does arise if it would be possible
to have in quantum systems a situation analogous to classical chaotic
structures?  In other words, if it would be possible for two slightly
different boundary conditions or some physical parameters
(for example, coupling constant) to correspond to two qualitatively
different wave functions?  Such dependence on physical conditions
in quantum systems may be analogous to classical chaotic dynamics.

As classical chaotic motion is more obvious in time discrete
systems which can be described, for example, by a discrete map (\cite{Hilborn} and
references therein), it is therefore only natural to study the quantum analogy of this
phenomenon by considering the quantum effects in solids which are
naturally discrete in space due to their atomic structure.   Consequently,
we study the  atomic lattice taking into account a nonlinearity
which arises due to electron-phonon interactions.  This interaction gives
rise to an apparent increase in the mass of the electron and causes the creation
of some self-trapped (or polaronic) states.  This phenomenon of the self-trapping
of electrons and excitons in solids, originally predicted theoretically
\cite{Peka,Rash1,Hols}, has been observed and well studied in many materials
(see, for example, review \cite{Rash2}).  It arises, primarily, in low
dimensional systems and in systems with a strong electron-phonon
interaction.  The (Rashba)--Holstein polaron model (\cite{Rash1,
Hols}) is described by a discrete nonlinear Schr\"odinger equation
with a $\phi^{4}$ type potential.  Other systems which may be
described by this generic equation include superconducting states
in layered superconductors, film deposition in Surface Science and
systems of coupled nonlinear oscillators.

\section{The Hamiltonian and the DNSE}

The classical and quantum systems mentioned above may be generally
described with the use of the Hamiltonian
\be
  H  = \sum_{i} | \psi_{i} - \psi_{i+1} |^{2} - \sum_{i} \frac{c}{2}|
  \psi_{i} |^{4} - E \sum_{i} \mid \psi_{i} \mid^{2},  \label{adpot1}
\ee
 where $\psi_i$ is (for the typical example) the wave
function of the self-trapped particle on the $i^{th}$ site, $c$ is
some (coupling) parameter and $E$ is the energy eigenvalue.  For coupled
(nonlinear) oscillators the function $\psi$ describes a lattice
distortion.  For quantum systems the wave function must, of course,
satisfy the conventional normalization condition
\begin{equation}
  \sum_i \mid \psi_{i} \mid^{2} = 1 \label{nc1}
\end{equation}
which is effectively described when the eigenvalue $E$ is used as a
Lagrange multiplier. The parameter $c$ may have both positive
(for  self-trapped quantum states) and negative (for nonlinear coupled
oscillators) values depending on the system being studied.

The energetically favourable states of a system described by
(\ref{adpot1}) correspond to the minima of (\ref{adpot1}). These
states are described by $\hat \nabla H(\psi) \equiv 0$, where
$\hat \nabla$ is the {\it differential operator} i.e.
$\frac{\partial H}{\partial \psi_{i}}$. Differentiating
(\ref{adpot1}) with respect to $\psi_{i}$ gives the conventional
discrete nonlinear Schr\"odinger equation (DNSE) of the form
\begin{equation}
   -\psi_{i-1}  + 2 \psi_i - \psi_{i+1} - c|\psi_i|^2 \psi_i  =
    E\psi_{i}  \label{dns1}
\end{equation}
which describes, amongst other things, the interaction of
electrons and phonons \cite{Rash1, Hols}.  $\psi_{i}$~is the wave
function of the polaron (electron localized by interactions with
phonons) so that $|\psi_{i}|^{2}$ is the charge density of the
polaronic state, $c$ is the electron-phonon coupling constant and
$E$ is the energy eigenvalue of this state. The normalization
condition then gives the charge of the electron. The electron
associated with a {\it small} polaron spends most of it's time
trapped on a single lattice site, that is, the majority of the
charge is distributed on the localizing site and (in the continuum
limit) the charge decays exponentially away from this lattice
site.

For simplicity, only real solutions are discussed
and so we make the substitution $|\psi_{i}|^2 = \psi_{i}^2$ in
the DNSE.  If we consider the discrete equation describing
nonlinear harmonic oscillators we have
\begin{equation}
   -x_{i-1}  + 2 x_{i} - x_{i+1} + c x_{i}^{3}  =  e x_{i}  \label{dns2}
\end{equation}
which differs from (\ref{dns1}) in the sign of $c$.
With the aid of the transformation
\be
  x_{n} = (-1)^{n} \psi_{n},
\ee
\be
e = 4 - E
\ee
we again get
\begin{equation}
   -\psi_{i-1}  + 2 \psi_{i} - \psi_{i+1} - c\psi_{i}^{3}  =
    E \psi_{i}.  \label{dns4}
\end{equation}

Thus, we find that all solutions and the classification of such solutions
for the quantum problem associated with the DNSE with positive coupling
constant $(c > 0)$  are also the solutions (and corresponding
classification) of the classical nonlinear lattices associated with
negative coupling constant $(c < 0)$.

Note that the behaviour of these solutions does not depend
explicitly on the parameter~$c$; $c$~can be rescaled or even
scaled out.  For example, for some scaling parameter $\beta$,
$\psi_{i}$ and $c$ are related by the similarity transformation:
$\psi_{i} \rightarrow \beta \psi_{i}$ and $c \rightarrow
c/\beta^2$ \cite{Kus-Rash}.  The normalization condition is then
given by $\sum\limits_{i} \psi_{i}^2 = 1 / \beta^2$.

\section{Iterated maps}

Conversion between Hamiltonian forms and mappings has long been known
to be a powerful mathematical tool for the theoretical and numerical
analysis of dynamical systems.  Indeed, two-dimensional maps allow
representation of a stationary configuration of a Hamiltonian of the form
(\ref{adpot1}) by a trajectory of a dynamical system.  Therefore, the
mathematical situation is {\it identical} to that of temporal evolution,
although the static problem has been formulated in terms of spatial
arrangements.  That is, we can apply adapted methods from the analysis
of chaotic dynamical systems and explore the nature of possible
solutions to the DNSE ie  we can represent the DNSE (\ref{dns4}) in the
form of a $2D$ map (\cite{KusK, Bak, Aubry}).

Introducing the auxiliary variable $Z_{i+1} = \psi_{i+1} - \psi_{i}$
gives
\begin{equation}
     Z_{i+1} = Z_i - E\psi_i - C\psi_i^3 \qquad {\rm and} \qquad
    \psi_{i+1} = \psi_i + Z_{i+1}
\label{2D-map}
\end{equation}
which may be iterated after choosing arbitrary initial
conditions.  Numerical experiments investigating different
trajectories of this $2D$ map have been performed for different
values of the parameters.  We find that there are essentially only
three types of phase portraits produced by iterating (\ref{2D-map}).
They can be {\it regular} (periodic) where only a small number of
points in the phase space are visited.  There are also {\it irregular
commensurate} (quasiperiodic) portraits in the form of closed loops.
These loops consist of a number of points being visited in the phase
space.  As the commensurability decreases the points visited in the
phase space become more and more dense and an elliptic orbit is mapped.
Finally, we have {\it irregular incommensurate} portraits where the
previous closed loops seem to be stochastically dispersed.

These phenomena, which are in agreement with previous work on the
subject (\cite{KusK, Bak, Aubry}), indicate that in
this system there are three qualitatively different types of
orbits which depend on the values of the parameters:
1) periodic, 2) quasiperiodic and 3) chaotic.  It is
well known that the regular/commensurate solutions
may be transformed into the quasiperiodic or chaotic type
structures by a slight change of initial conditions or parameters.

Below we discuss an alternative method for solving (\ref{dns4})
and present the spectrum of~(\ref{dns4}) in the large $c$ limit.
We then go on to apply a numerical method to our system of DNSEs
and obtain results which we cannot reproduce with the above
mapping procedure. These results are contrasted with those
obtained using the iterated map technique and the spatial
structure of the wave function is examined.

\section{Exact solutions in the limit $\pbf{c \rightarrow \infty}$}

The above DNSE (\ref{dns4}) has exact
solutions in the limit $c \rightarrow \infty$ for a $N$-site lattice,
which have been published separately \cite{KusD}.
The associated energy eigenvalue of these exact solutions takes the form
(for $c \gg 1$)
\begin{equation}
    E = \frac{2m + 4l - c}{n},    \label{eieq}
\end{equation}
where the eigenvalue $E$ intrinsically depends on both the
structure of the wave function on the lattice --- described by
$m$, $l$, $n$ --- and on the parameter $c$. The structure of the
wave function is such that it is localized, $\psi_{i} \neq 0$, on
just $n$ sites of the $N$-site lattice ($n \leq N$); on the
remaining $N - n$ sites the wave function is zero, $\psi_{i}= 0$.
Note that on each site where the (normalized) wave function is
localized it takes a value of either $\psi_{i} = \pm 1/ \sqrt{n}$
for all $n$ sites --- no other values are taken. These $n$
localizing sites are separated into $m$ spots --- groups of
neighbouring lattice sites on which the wave function is localized
--- and $l$ kinks --- change in the sign of the wave function on
adjacent lattice sites --- inside the spots. Between the spots the
wave function is vanishing. Both $m$ and $l$ are less than or
equal to $n$.

Eq. (\ref{eieq}), obtained in the limit $c \rightarrow \infty $,
has been compared with the exact and numerical
solutions for various systems consisting of different numbers of sites.
In all these cases for nearly all values of $c$ (except small regions
of critical values where the self-trapped solutions originate)
there is perfect agreement with the derived formula (\ref{eieq}).
However, in contrast with this perfect agreement between
eigenvalues, a decrease in the value of $c$ leads to a noticeable
deviation in the wave functions (eigenvectors)  from those
obtained in the limit $c\rightarrow\infty$.

The form of (\ref{eieq}) suggests that as well as the ground state of
the polaron $(E = 2-c)$ where the polaron is trapped on a
single lattice site there also exist excited trapped states, for
example $E = (2-c)/2$, where the charge density of the polaron is
distributed over~2 sites adjacent to one another.  This is in contrast
with previous work on this subject where the continuum approximation is
made.  We find that the continuum approximation is insufficient because
it only describes the ground state of the system; there exist a whole
host of excited states in a discrete system which simply do not exist in
the continuum limit!

\section{Exact numerical solutions for finite $\pbf{c}$}

The first order corrections to the wave function (eigenvectors) obtained with
the use of perturbation theory are of the order of $\sqrt{n}/c$ ie
$O(1/c)$.  When the coupling constant $c$ is not very large
the wave functions of some states have interesting incommensurate and
chaotic structures.  Thus, from comparison with numerical results
and  with perturbation theory we conclude,
that even though the spectrum of the DNSE for a system
with a finite, arbitrary number, $N$, of sites
is well described by equation (\ref{eieq}) for $c \gg 1$,
the shape of the appropriate
wave function for smaller values of $c$ may have
only qualitative features of the wave function
obtained in the limit $c \rightarrow \infty$.
As $c$ decreases the localization spots smear out.
Since the spectrum, (\ref{eieq}), is associated with a local
localization pattern, it is universal and
does not depend on the boundary conditions.

Although the structure of the wave function for $c \gg 1$ does not
strongly agree with the structure of the wave function in the
limit $c \rightarrow \infty$ we find that the wave function may be
approximated reasonably well by an exponential function.  This is
only valid if the peaks in the wave function structure are
separated sufficiently so that there is little or no interaction
between the tails.  For lattice sites sufficiently far away from
the localized site we can assume that in (\ref{dns4}) the value of
the wave function is small so that the cubic nonlinear term is
negligible, $\psi_{i}^{3} \approx 0$.  Then, in the continuum
limit, we obtain a second order linear differential equation which
may be expressed as
\begin{equation}
  -\frac{\partial^2 \psi}{{\partial x^2}} = E \psi(x).  \label{conteq1}
\end{equation}

(\ref{conteq1}) has the asymptotic solution
\begin{equation}
  \psi = A \exp\left(- \sqrt{-E}x\right),  \label{conteq2}
\end{equation}
 where $A$ is some parameter which from the normalisation
condition goes as $A \sim (-E)^{1/4}$.   (\ref{conteq1}) may only
be used to describe the behaviour of the wave function
sufficiently far away from the local maxima of the wave function
because near the maxima the cubic nonlinear term cannot be
neglected.  Hence, away from the peak the behaviour of the wave
function may be described as an exponential decay.

If a small perturbation in the asymptotic limit ($c \rightarrow
\infty$) wave function ${\pbf \Psi} = (\psi_{1}, \psi_{2},
\ldots$, $\psi_{i}, \ldots, \psi_{N})$ and energy eigenvalue $E$
is considered then a correction to the wave function is revealed.
This correction is found by starting with the asymptotic solution
of the DNSE {\it but taking finite $c$} and successively applying
the Newton--Raphson method until the corrections to the wave
function become negligible and ${\pbf \Psi}$ converges to some
limit.

In matrix notation the $N$-dimensional Newton--Raphson method takes
the form
\begin{equation}
   {\pbf \Psi} (k+1) = {\pbf \Psi}(k) - \left[ \hat \nabla^{2} H ({\pbf \Psi}
   (k)) \right]^{-1} \hat \nabla H ({\pbf \Psi} (k)) , \label{dnr}
\end{equation}
 where ${\pbf \Psi}(k)$ is our wave function (vector) consisting of
$N$ components ($N$ lattice sites) iterated $k$ times --- if $k=0$
then it is simply the wave function in the asymptotic limit~---
and $\hat \nabla^{2} H ({\pbf \Psi} (k))$ is (for 3 or more sites)
a tridiagonal $N\times N$ matrix with diagonal elements $2 - E - 3
c \psi_{i}^{2}(k)$, elements on either side of the diagonal $= -1$
and top-right and bottom-left elements $= -1$ for periodic
boundary conditions (PBC).

This procedure is used until the correction to the wave function
becomes negligible.  This means that the wave function converges
to some limit.  This limit is an exact, numerical solution of the
DNSE (\ref{dns4}).  Note that, as the corrections are of
the form of a series with negative integral powers of $c$ (ie
first order correction $\sim 1/c$, second order correction
$\sim 1/c^{2}$, etc), in the limit $c \rightarrow \infty$ the
corrections are all zero and the asymptotic solutions are obtained.

The energy eigenvalue (for PBC) after each iteration step is given by
\begin{equation}
  E(k) = \frac {-c \sum\limits_{i} \psi_{i}^{3}(k)} {\sum\limits_{i} \psi_{i}(k)}
  \label{enrgycorrm}
\end{equation}
and the value of $E(k)$ converges towards $E$
for most values of $c \gg 1$ except for small regions of $c$ where the
localised states originate.  Even in these sensitive regions the value
of $E(k)$ is in fairly close agreement with $E$.

This process is, however, not always convergent.  The wave vector may
change structure when different eigenenergies have the same value
or when the structure obtained from the asymptotic limit does not
exist for a range of $c$.  For example, for a 3 site system with PBC
two of the eigenvalues are $-c/3$ and $(6-c)/2$.  Both of these eigenvalues
have the same numerical value for $c=18$ but they have very different structures
of the wave function.  It is possible for
the initial wave vector structure corresponding to one energy eigenvalue
to change during the iteration process to another structure which corresponds
to a different energy eigenvalue.  However, this new structure does also
exist and is also a solution of the DNSE (\ref{dns4}).

If during the numerical process there is a
drastic change in $E$ then this indicates that the wave
function structure has changed and the iteration process will converge to
some other limit.

Applying this method to the above DNSE for finite $c$ taking
the asymptotic solutions (which are exact in the limit
$c \rightarrow \infty$) as the initial condition we may compare the
results obtained with Newton's method with the results
obtained by the use of $2D$ iterated maps.
For such a comparison, however, we have to introduce the new {\it phase
space function} $\psi_{i+1} - \psi_{i}$ for the Newton procedure applied to
the discrete system of equations and
plot $\psi_{i+1} - \psi_{i}$ against $\psi_{i}$. This then allows us
to compare the wave function structures obtained with the use of the
Newton procedure and the trajectories of the $2D$ iterated maps
discussed in Section 3.  We find that as well as obtaining results from
the two different methods which are consistent with each other we
also obtain results from the Newton procedure which we could not get
with the $2D$ maps.  That is the pictures representing the phase
space of our quantum system correspond to different periodic or chaotic
spatial structures of the wave function.

\section{Results}

In the previous section we have reformulated our problem of the
solution of a discrete nonlinear set of equations, (\ref{dns4}),
to a (discrete Newton--Raphson) iterative problem.  This gives
possible criteria for the classification of the solutions obtained.
Following this iteration procedure we find that there exist different
types of structures of the wave function; nonlinear
localized solutions arise for $E<0$.
Our results indicate that there exist both regular and irregular
types of structure of the wave vector for negative values of $E$.
These structures may be akin to periodic, quasiperiodic and
(deterministic) chaotic structures.

Of course for finite systems quasiperiodic and chaotic structures are not
well defined because these are, strictly speaking, well defined only for
systems of infinite size.  However, we may still indicate
analogous features, for example, with the aid of phase portraits
adopted for finite size systems which are possibly
equivalent to regular commensurate, irregular commensurate and irregular
incommensurate structures of the wave function on the lattice.

\subsection{Periodic trajectories}

Fig.~1 shows the final (converged) behaviour of the wave function
on the lattice (1a) and the corresponding phase portrait (1b)
obtained by the Newton method described above for a 100 site
lattice with Periodic Boundary Conditions (PBC). The initial
structure consisted of a normalized wave function being localized
on only 10 sites all equally spaced apart. Each of these localized
sites is separated by 9 sites where the wave function is zero. The
sign of the wave function alternated ie if it is positive at a
particular site then the sign of the wave function 10 sites before
or 10 sites after is negative. This gives rise to a period 20
structure on a 100 site lattice. The corresponding energy
eigenvalue (given by (\ref{eieq})) is $E = (20 - c)/10$. Taking
the coupling constant to be $c = 24$ and applying the above
procedure to this initial structure of the normalized wave
function we obtain the solution presented in Fig.~1a.  This shows
a periodic behaviour of the wave function through the lattice (as
expected from the initial conditions).  The corresponding phase
portrait of the quantum state, Fig.~1b, indicates that this
initial structure gives rise to a~regular commensurate period 20
structure. The phase portrait consists of only 20 points as the
wave function oscillates regularly.

\begin{figure}[th]

\centerline{\epsfig{file=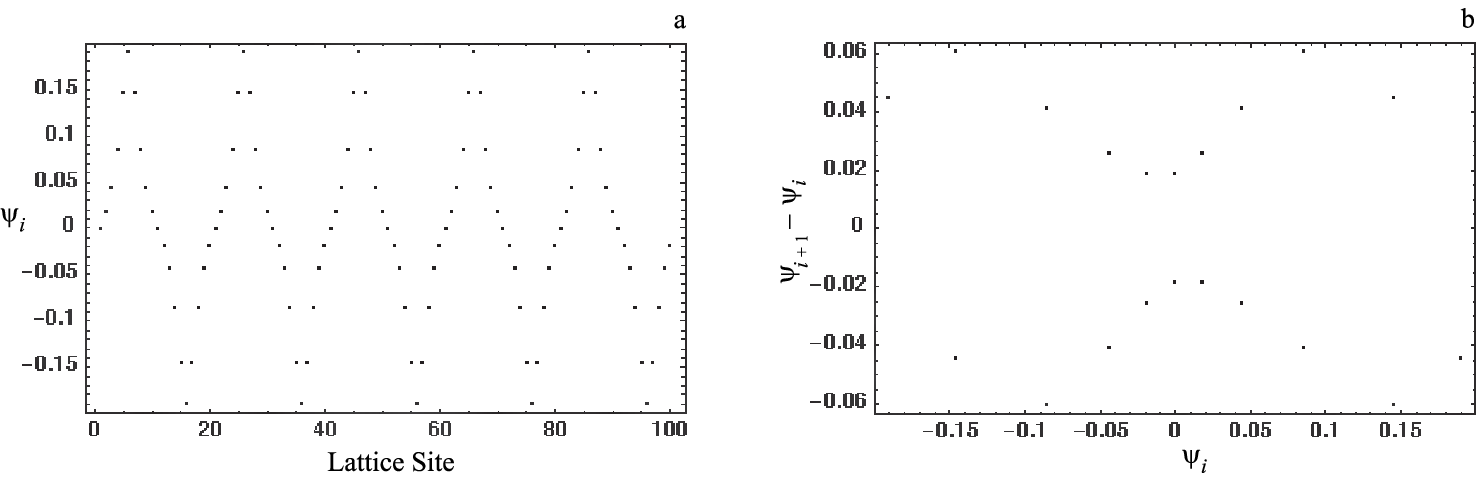,width=150mm}}

\vspace{-3mm}

\caption{}
\end{figure}

\begin{figure}[th]

\centerline{\epsfig{file=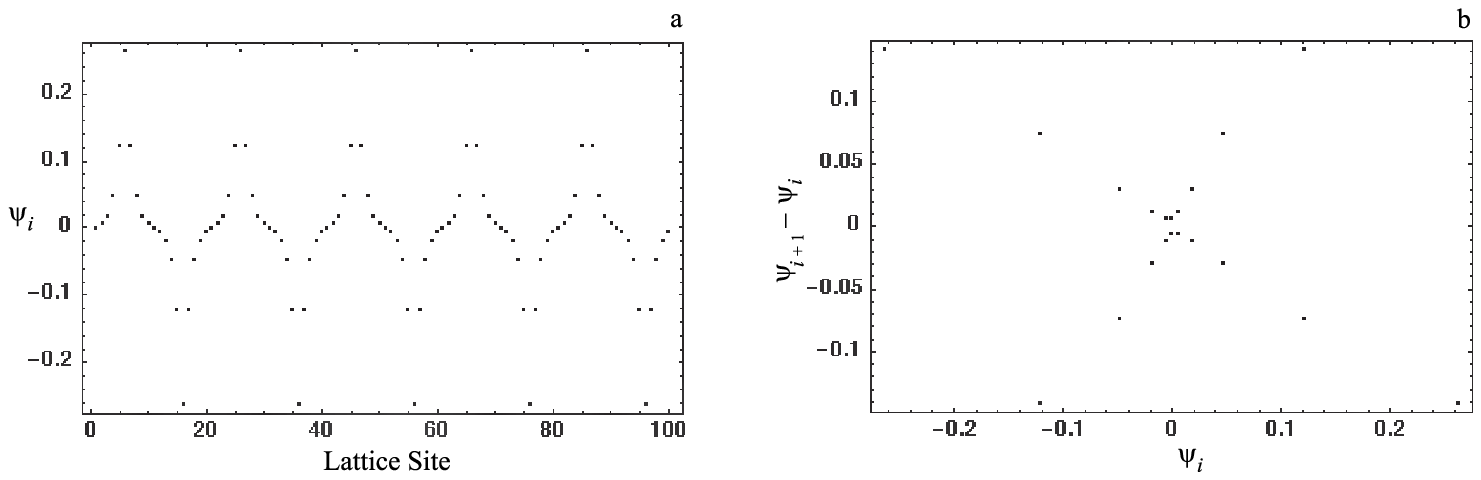,width=150mm}}

\vspace{-3mm}

\caption{}

\vspace{4mm}


\centerline{\epsfig{file=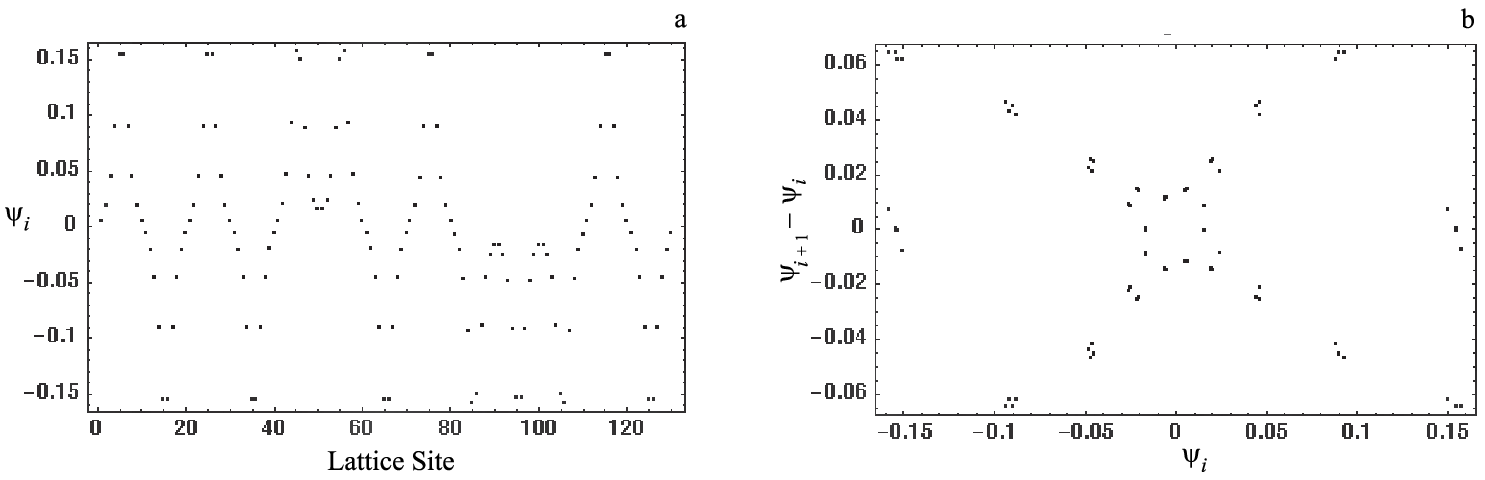,width=150mm}}

\vspace{-3mm}

\caption{}

\vspace{-2mm}
\end{figure}

Fig.~2 shows what happens as $c$ is increased to $c = 30$.  The
width of each peak (or trough) decreases and the amplitude
increases, that is, the wave function becomes more localized.  In
the limit $c \rightarrow \infty$ we will obtain our initial
structure of the wave function being completely localized on the
original 10 sites and zero everywhere else.  Note that in this
limit the corresponding phase space will consist of just 5 points:
the origin $(\psi_{i}, \psi_{i+1} - \psi_{i}) = (0, 0)$,
$(\psi_{i}, \psi_{i+1} - \psi_{i}) = \left(0, 1/ \sqrt{n}\right)$,
$(\psi_{i}, \psi_{i+1} - \psi_{i}) = \left(0, -1/
\sqrt{n}\right)$, $(\psi_{i}, \psi_{i+1} - \psi_{i}) =\left(1/
\sqrt{n}, -1/ \sqrt{n}\right)$ and $(\psi_{i}, \psi_{i+1} -
\psi_{i}) = \left(-1/ \sqrt{n}, 1/ \sqrt{n}\right)$.

\subsection{Irregular trajectories}

Fig.~3 shows the behaviour of the wave function on the lattice
(3a) and the phase portrait~(3b) for a 130 site lattice with PBC.
The initial structure consisted of a normalized wave function
being localized on 26 sites separated into 13 spots, each of which
consisted of two localizing sites with the same sign of wave
function.  These spots were equally spaced apart and separated
from each other by 8 sites where the wave function is zero.
However, the sign of the wave function of different spots did not
change regularly.  The corresponding energy eigenvalue is $E = (26
- c)/26$. Taking the coupling constant to be $c = 40$ and applying
the above procedure to this initial structure of the normalized
wave function we obtain the solution presented in Fig.~3a.  This
shows an irregular oscillation of the wave function through the
lattice although each of the peaks (and troughs) of the wave
function are similar to one another.  The corresponding phase
portrait, Fig.~3b, maps out a double loop and near the origin is
reminiscent of a hyperbolic point.

Again as $c$ is increased the wave function becomes more localized
and in the limit $c \rightarrow \infty$ we again obtain our
initial structure of the wave function being completely localized
on the original 26 sites and zero everywhere else.  Note that in
this limit the corresponding phase space will consist of 7 points:
the five listed above and two more: $(\psi_{i}, \psi_{i+1} -
\psi_{i}) = (1/ \sqrt{n}, 0)$ and $(\psi_{i}, \psi_{i+1} -
\psi_{i}) = \left(-1/ \sqrt{n}, 0\right)$.

\subsection{Random trajectories}

Fig.~4 shows the behaviour of the wave function on a 208 site
lattice for an arbitrary initial configuration.  The energy
eigenvalue given by (\ref{eieq}) is $E = (86 - c)/63$.  Taking $c
= 260$ we obtain Fig.~4a where the behaviour of the wave function
appears disordered. The wave function structure has split into
three regions: top, middle and bottom.  Fig.~4b shows the
underlying pattern manifest in the wave function which consists of
narrow peaks and troughs arranged in a pattern without any
long-range order.  The corresponding phase space, Fig.~4c, is in
the form of a dispersed closed loop. This structure (which is
reminiscent of chaos in classical systems) is due to the
interaction of different localization spots.  The appearance of a
stochastic orbit indicates the effects of irregular
incommensurability of the wave function on the lattice.

\begin{figure}[th]
\vspace{-1mm}

\centerline{\epsfig{file=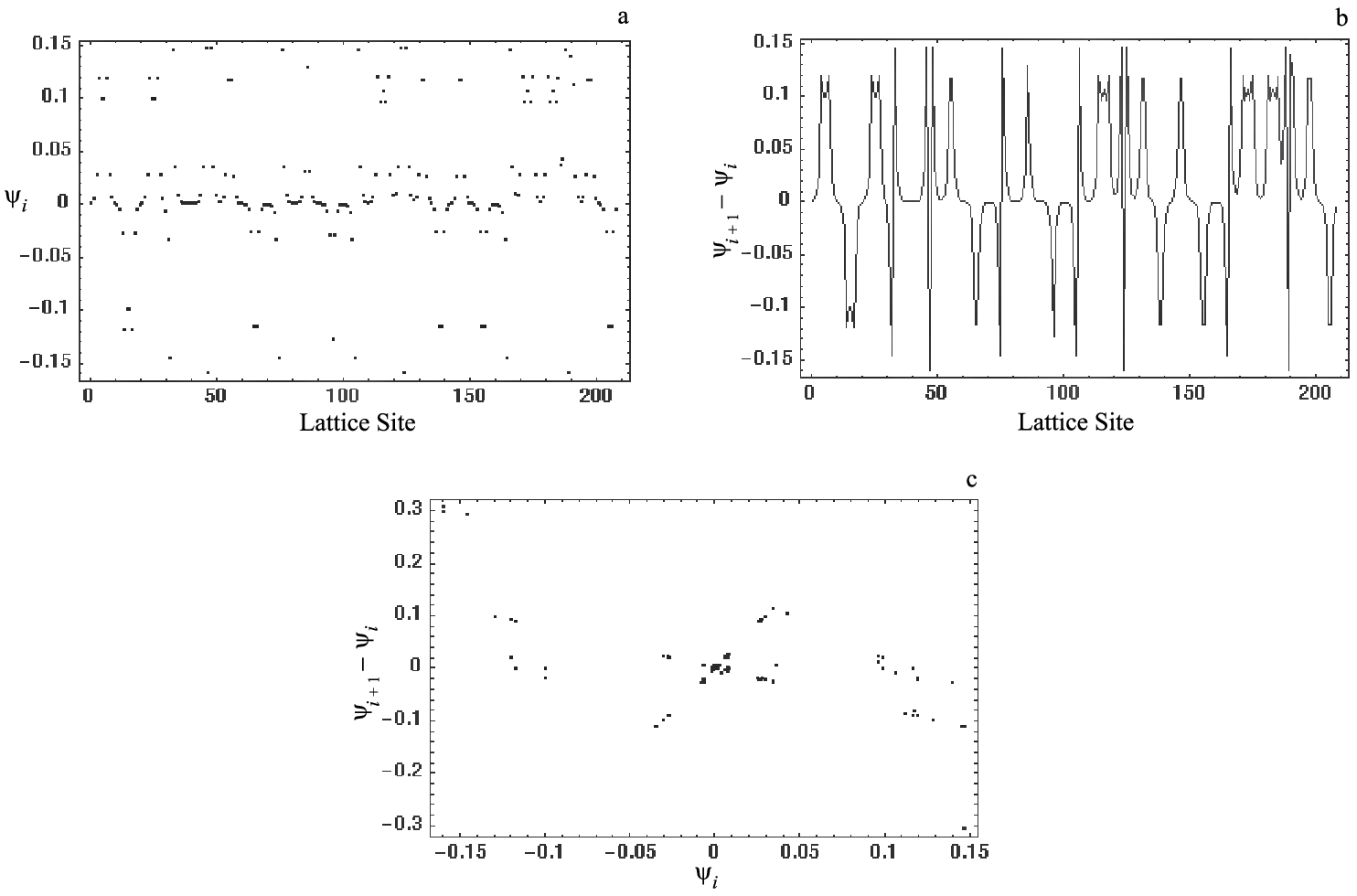,width=150mm}}

\vspace{-1mm}

\vspace{-3mm}

\caption{}

\end{figure}

\begin{figure}[th]
\vspace*{-1mm}

\centerline{\epsfig{file=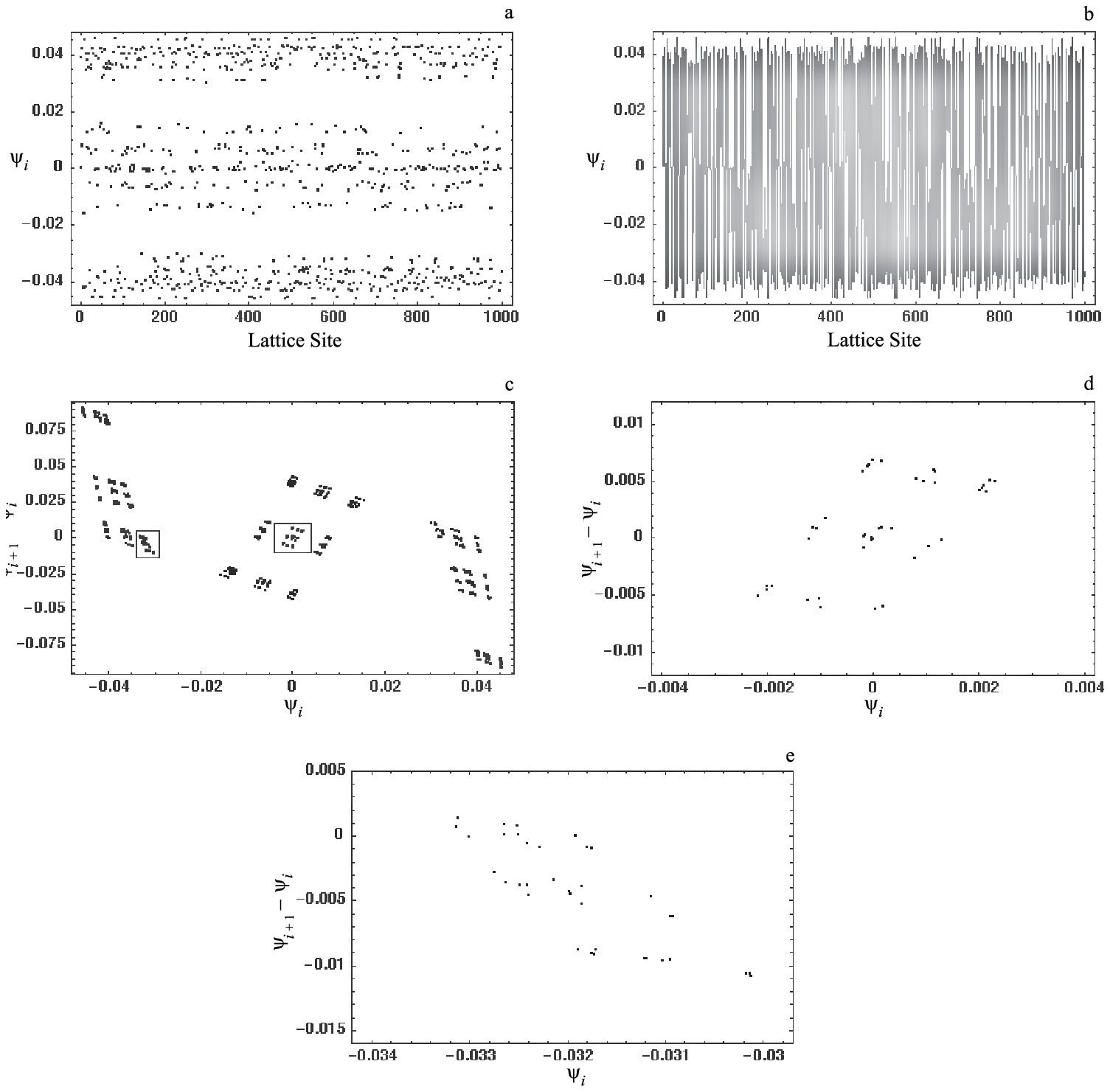,width=150mm}}
\vspace{-3mm}
\caption{}
\vspace{-3mm}
\end{figure}

Fig.~5 shows the behaviour of the wave function on a 1000 site
lattice.  The initial confi\-gu\-ration was generated (pseudo-)
randomly from the integers $-1, 0, 1$ and then normalized.  The
energy eigenvalue (from (\ref{eieq})) is $E = (1238 - c)/636$ and
a value of $c = 4000$ gives the wave function structure shown in
Fig.~5a.  Again the wave function structure has split into top,
middle and bottom.  Fig.~5b shows the (dense) oscillations of the
wave function and Fig.~5c is the phase portrait for this wave
function.  The points in this phase portrait don't seem to lie on
an obvious closed loop or even a dispersed closed loop but rather
a~smeared curve.  As this structure has no definate period it
could be a possible candidate for a chaotic state of the system.
The phase portrait is split into three parts: top-left, centre and
bottom-right.  Each of these three sections is further divided
into sub-sections in which there appears a ``nesting'' of points.
Something which is especially noticeable in the central part of
this phase portrait is that each of the sub-sections are similar
to the larger section in which they are contained.

For example, the boxed
sub-section of nested points at the origin, which is magnified in
Fig.~5d, is similar to the central section of Fig.~5c.  The same
trend can also be observed in other parts of the phase portrait, for
example Fig.~5e is a magnification of the boxed region in the
top-left segment of Fig.~5c.  Both of these examples indicate a
self-similarity in the structure of the wave function on a lattice.
That is similar structures of the wave function exist on different
length scales which suggests a fractal nature of the wave function on
the lattice.

As the value of $c$ is increased the different
points visited in the phase space converge until only a maximum of
nine points exist.  These nine points which correspond to the wave
function structure in the limit $c \rightarrow \infty$ are
$(\psi_{i}, \psi_{i+1} - \psi_{i}) = (0, 0)$,
$(\psi_{i}, \psi_{i+1} - \psi_{i}) = \left(0, 1/ \sqrt{n}\right)$,
$(\psi_{i}, \psi_{i+1} - \psi_{i}) = \left(0, -1/ \sqrt{n}\right)$,
$(\psi_{i}, \psi_{i+1} - \psi_{i}) = \left(1/ \sqrt{n}, 0\right)$,
$(\psi_{i}, \psi_{i+1} - \psi_{i}) = \left(1/ \sqrt{n}, -1/ \sqrt{n}\right)$,
$(\psi_{i}, \psi_{i+1} - \psi_{i}) = \left(1/ \sqrt{n}, -2/ \sqrt{n}\right)$,
$(\psi_{i}, \psi_{i+1} - \psi_{i}) = \left(-1/ \sqrt{n}, 0\right)$,
$(\psi_{i}, \psi_{i+1} - \psi_{i}) = \left(-1/ \sqrt{n}, 1/ \sqrt{n}\right)$,
$(\psi_{i}, \psi_{i+1} - \psi_{i}) = \left(-1/ \sqrt{n}, 2/ \sqrt{n}\right)$.

\section{Summary}

Thus, we obtain that in the limit $c \rightarrow \infty$ there
arises a degenerate set of solutions associated with different
localization states of the polaronic wave function in the
(Rashba)--Holstein model which consist of empty lattice sites
(where the wave function is vanishing) and lattice sites where the
wave function is localized. These localization patterns, which can
be viewed as soliton type structures, have many different
configurations which correspond to the same eigenvalue of the
DNSE.  However, when the value of $c$ is not infinite this
degeneracy is broken.  Different localization spots within the
pattern start to interfere with each other and modify the
behaviour of the wave function. This leads to the existence of
excited trapped states in this discrete system as well as the
ground state.  These excited trapped states may be experimenatlly
observed by imaging the local density of states (LDOS).  A finite
amplitude of the LDOS fluctuations manifests correlations between
fluctuations of local densities of individual wave functions at
close energies.  We go on to apply the Newton--Raphson method to
this system of discrete nonlinear equations where, starting with
the asymptotic solutions and iterating, ${\pbf \Psi} (k+1)$ is in
general a better approximation to the solution of (\ref{dns4})
than ${\pbf \Psi} (k)$ for finite $c$.  This solution gives the
structure of the wave function on the lattice and suggests the
possible existance of fractal structures where the wave function
structure at one scale is echoed at another scale.

The interaction between the solitons is essentially governed by the radius of
the solitons and their separation and depends on the reciprocal of the
modulus of an eigenvalue of the DNSE,
$1/ \sqrt{|E|}$, which governs the radius of the soliton peaks.
If the radius of the soliton peaks is much smaller than the separation
then we have regular, commensurate behaviour of the polaronic wave function
on the lattice.  In this case the interaction between the solitons is very weak.
As the radius of the solitons increases
($1/ \sqrt{|E|}$ increases) and the separation remains constant the
interaction between the solitons increases and the behaviour of the
wave function on the lattice becomes less regular.  Finally, when the
radius of the soliton peaks is about the same as the separation of the solitons
we have strong interactions between the solitons.   This results in
strong incommensurability of the behaviour of the wave function on
the lattice.

We find that the interference between the spots can give rise to
three qualitatively different structures: periodic, quasiperiodic and
chaotic.  To indicate such structures in the quantum state the methods
used in the studies of classical chaos were applied to the quantum
system.  A phase portrait was built up which
consists of the amplitudes of the wave functions and of residues
of these amplitudes associated with neighbouring sites.  That is it
is a projection of the Hilbert  ``phase space of our quantum system''
in the plane, that is, the set $\{\psi_{i+1} - \psi_{i}, \psi_{i}\}$.

\newpage

In the regular periodic and quasiperiodic solutions the
wave function amplitudes replicate with some period equal to some
integer number of lattice constants or creates aperiodic structures,
respectively.
However, there is also the appearance of structures analogous to the
those arising in classical chaos which gives rise to self-similar
structures of the wave function on the lattice at different length
scales.
The destruction of the periodic and quasiperiodic
orbits, which has been ascribed to the creation of chaotic structure
in the wave function, is also exhibited by the system.  That is the
creation of a structure which has no definite period has been found.

\subsection*{Acknowledgements}

KEK gratefully acknowledges helpful assistance by R~Valder
at the computer center of the Heinrich-Heine University
in D\"usseldorf, by J~Fertillet, J~C~Fiers and
M~Lams at the Universit\'e du Littoral in Dunkerque.

\label{dhillon-lastpage}

\newpage

\end{document}